\documentclass[aps,prl,onecolumn,superscriptaddress,groupedaddress]{revtex4}  
\usepackage{graphicx}  
\usepackage{dcolumn}   
\usepackage{bm}        
\usepackage{amssymb}   
\usepackage{amsmath}
\usepackage{xcolor}
\begin{document}

\title{Aging of Thermoreversible Gel of Associating Polymers}

\author{ Kulveer Singh }
\email{kulveersingh85@gmail.com}
\author{Yitzhak Rabin}%
 \email{yitzhak.rabin@biu.ac.il}
\affiliation{Department of Physics, and Institute of Nanotechnology and Advanced Materials, \\Bar-Ilan University,
Ramat Gan 52900, Israel}%

\date{\today}

\begin{abstract}
We performed Langevin dynamics simulations of amphiphilic polymers that contain strongly associating stickers connected by long soluble chain segments.  We explored the aging of a thermoreversible gel which is crosslinked by aggregates (clusters) of the stickers and observed a dramatic increase of their lifetime with aging. We find that the observed aging is the result of structural reorganization of the network: the number of crosslinks decreases with time as small clusters disintegrate and their stickers are reabsorbed by large clusters that grow and become more stable by forming additional intramolecular associations between their stickers. 
\begin{figure}[h!]
\text{TOC figure}\par
\includegraphics[width=0.4\linewidth]{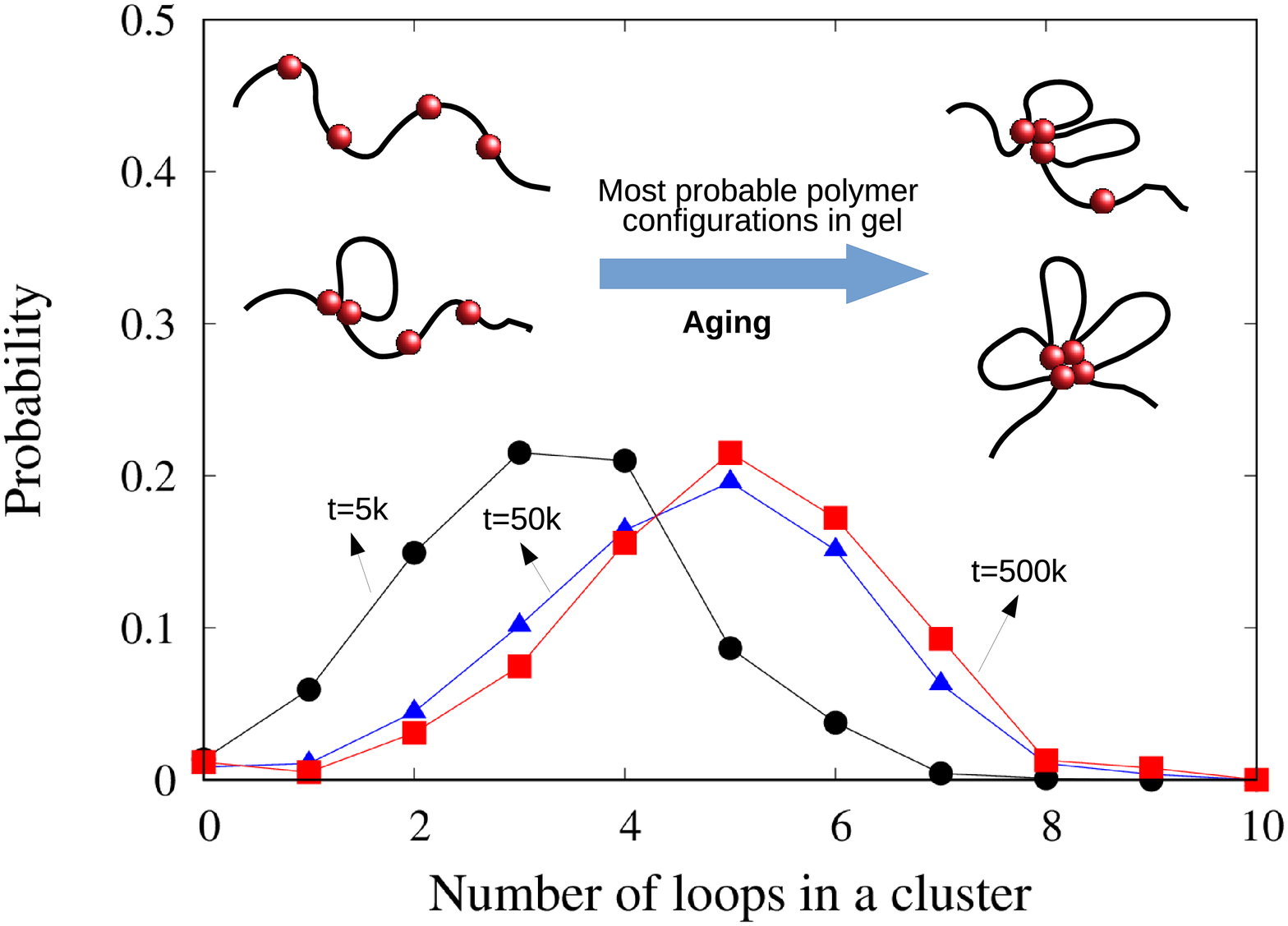}
\end{figure}

\end{abstract}

\maketitle

\section{Introduction}
Associating polymers are heteropolymers which contain segments that attract each other via non-covalent interactions such as hydrogen bonding, hydrophobic and dipolar interactions, etc. \cite{Strauss1989, Chassenieux2011, Colby}. In solution, these attractive segments can form intramolecular associations with similar segments of the same polymer or intermolecular associations between different polymers, depending on polymer concentration. Theoretical studies show that at low concentrations, intramolecular associations are favored leading to the formation of loops \cite{Cates1986} and flower-like micelles \cite{Borisov1995},  whereas at high concentrations intermolecular associations dominate and result in the formation of a macroscopic thermoreversible network in solution (in other regions of the phase diagram gelation may be replaced by phase separation) \cite{Semenov1995, Winnik1997, Semenov1, Semenov2, Rubinstein1999, Dobrynin2004}. These physical gels exhibit viscoelastic behavior and can be classified as liquidlike or solidlike depending on the stability of the interchain associations that act as physical crosslinks. When the lifetime of these effective crosslinks is relatively short (on experimentally relevant time scales) the network behaves as polymer fluid with $G'\ll G''$ where $G'$ and $G''$ are the storage and loss moduli, respectively. Conversely, in the limit of large crosslink lifetimes (again, compared to experimental time scales), these gels exhibit solidlike behavior ($G'\gg G''$) \cite{Rubinstein}. The structure of the gel depends on the number of attractive segments, on the strength of the interaction between them, and on their distribution along the polymer backbone.     

Many physical gels show aging behavior \cite{Gomez2013, Piazza2013}. For example, in an experimental study of methylcellulose gel (a thermoreversible network formed by aggregation of hydrophobic chain segments at elevated temperature), very slow reorganization of the network following its formation by temperature quench was observed by multispeckle dynamic light scattering, and it was shown that the timescale of density variations on length scales comparable to the wavelength of light, increases with the age of the gel \cite{Rabin}. In this paper we use computer simulations of a simple model of associating polymers in implicit solvent \cite{Dino} to study the aging of a thermoreversible gel. Similarly to  ref. \cite{Semenov1995}  where networks made of telechelic polymers were studied theoretically using mean field methods, we find that such gels are crosslinked via the formation of clusters of attractive segments  (stickers), connected by soluble polymer segments. When the attractive interactions between stickers are sufficiently strong, we observe a dramatic increase of the lifetime of the crosslinks with aging time. We find that the observed aging is the result of structural  reorganization of the clusters whose size increases with time and then saturates. Somewhat counterintuitively, we find that this reorganization takes place at fixed average number of polymers per cluster and is the consequence of the fact that the number of {\it{intramolecular}} associations and therefore of loops, increases with time.

\section{Model and Methods} 
We consider a system of $M$  polymers, of $N=50$ segments (beads) each. The backbone of the chain is connected by finitely extensible nonlinear elastic (FENE) bonds, with the potential
\begin{equation}
	U^{FENE}=-0.5KR_0^2ln \left[ 1-\left(\frac{r}{R_0}\right)^2 \right] + 4\epsilon\left[\left(\frac{\sigma}{r}\right)^{12} - \left(\frac{\sigma}{r}\right)^{6}\right] + \epsilon 
\label{eq:fene},
\end{equation}
where we take $K=30.0$, $R_0=1.5$, $\epsilon=1.0$, $\sigma = 1.0$ and truncate the $2^{nd}$ term on the rhs of the above equation at  $2^{1/6}\sigma$ (thus taking only the repulsive part of the Lennard-Jones potential). Beads which do not have a FENE bond between them interact via the Lennard-Jones potential 
\begin{equation}
	U^{LJ}_{ij}(r)=4\epsilon_{ij}\left[\left(\frac{\sigma}{r}\right)^{12}-\left(\frac{\sigma}{r}\right)^{6}\right] \hspace{1cm} r<r^{cut}_{ij}
\label{eq:LJ}.
\end{equation}

Here $U^{LJ}_{ij}(r)$ is truncated and shifted to zero at cutoff distance $r^{cut}_{ij}$. Each polymer contains two types of beads designated as stickers and non-stickers respectively, such that $\epsilon_{ij} = \epsilon_s$ if $i^{th}$ and $j^{th}$ beads are stickers and  $\epsilon_{ij} = \epsilon_{ns}$ if at least one of $i^{th}$ and $j^{th}$ beads is a non-sticker ($r^{cut}_s$ and  $r^{cut}_{ns}$ are the corresponding cutoff distances). We use LAMMPS \cite{Plimpton1995} (Large-scale Atomic/Molecular Massively Parallel Simulator) to carry out Langevin dynamics simulations in the NVT ensemble. The simulation is performed in a box of size $51\times 51\times 51$ in units of $\sigma$, using periodic boundary conditions (the box size was chosen to be slightly bigger than the stretched length of the polymer). The  motion of each bead is given by the Langevin equation, neglecting hydrodynamic interactions
\begin{equation}
m\ddot{\bf{r}}_i(t) = -\frac{\partial U}{\partial {\bf{r}}_i} -\zeta \dot{\bf{r}}_i(t) + \eta_i(t)
\label{eq:langevin},
\end{equation}
where $U$ (sum over all $U_{ij}$),  $\zeta$ and $\eta_i$ are the total potential energy, bead friction coefficient and random thermal force due to implicit solvent, respectively.  The rms amplitude of the random noise is proportional to  $(\zeta k_BT/\Delta t)^{1/2}$, where $k_B$, $T$ and $\Delta t$ are Boltzmann's constant, temperature and integration time step, respectively. All the time scales in the following are expressed in Lennard-Jones time units $\tau_{LJ} = (m\sigma^2/\epsilon)^{1/2}=1$ (the mass $m$, particle diameter $\sigma$, interaction parameter $\epsilon$, and $k_B$ are all set to 1). 

We set the integration time-step to be $\Delta t =0.005$, friction coefficient $\zeta = 0.02$, and fixed the temperature at $T = 1$. Following ref. \cite{Dino} we took $4$ stickers per polymer which are separated by $9$ non-sticker beads along the chain (the two non-sticker end segments are of lengths $9$ and $10$, respectively). To obtain an initially uniform polymer solution, we placed all the polymers in an array inside the simulation box. We assigned $\epsilon_{ns}= \epsilon_{s}=1.0$ and $r^{cut}_{ns}=r^{cut}_{s}=1.12\sigma$ (since the potential is shifted to zero at the cutoff, this yields purely repulsive interactions between all beads) and equilibrated the polymer solution for sufficiently long time to reach equilibrium. We then changed the interaction parameter between stickers to $\epsilon_{s}=6.0$ and the corresponding cutoff distance to $r^{cut}_{s}=2.5\sigma$ and continued to monitor the system through the processes of gel formation and aging.

\section{Results}
Since our goal is to study the aging of gels, we began by choosing the polymer concentration well-above the sol-gel transition (we fixed the temperature at $T=1$ and, therefore, polymer concentration is our tunable parameter). The extent of gelation can be quantified by the parameter $M_{LC}/M$, where $M_{LC}$  is the number of polymers in the largest connected aggregate in the system \cite{Dino}. In order to define connectivity, we defined an association (bond) between two stickers if they are within a range of $1.5\sigma$ from each other; two polymers are considered to be crosslinked if they have at least one bond between them. A fully connected system corresponds to $M_{LC}/M=1$ where all the polymers belong to a single aggregate, whereas $M_{LC}/M=0$ corresponds to a perfect sol state where none of the polymers in the solution are bound. 

Starting from a random initial condition (obtained by equilibrating the system under good solvent conditions for the polymers), we performed simulations for different volume fractions, $\phi$, and computed $M_{LC}/M$ as function of time for each $\phi$. We found that the size of the largest cluster rapidly increases with time and eventually saturates at a plateau value of $M_{LC}/M$ (a plot of the steady state value of $M_{LC}/M$ vs $\phi$ is shown in fig. S1A in SI). Since we are interested in the case where almost all the polymers belong to the network, we chose the polymer volume fraction $\phi=0.096$ for which approximately $98\%$ of the polymers belong to the connected network. This volume fraction exceeds the overlap volume fraction of polymers in good solvent (with $\epsilon_s=\epsilon_{ns}=1$) which we determined to be $\phi^{\ast}=0.059$ (not shown). Once the attractive interaction between the stickers is switched on, the network forms very fast and $M_{LC}/M$ reaches saturation at $t \approx 100\tau_{LJ}$ (see fig. S1B in SI). 
\begin{figure}[h]
\includegraphics[width=0.5\linewidth]{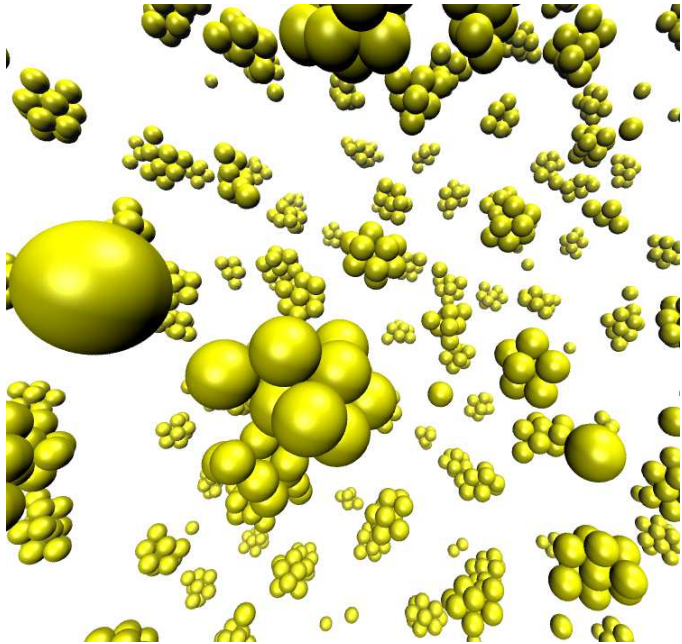}
	\caption{ \label{fig.snapshote6} Snapshot of the clusters of stickers in the system at time $t=500k$ ($5\cdot 10^5 \tau_{LJ}$). Polymer segments that consist of non-stickers are not shown.}
\end{figure}
 
\begin{figure*}[ht]
\includegraphics[width=\linewidth]{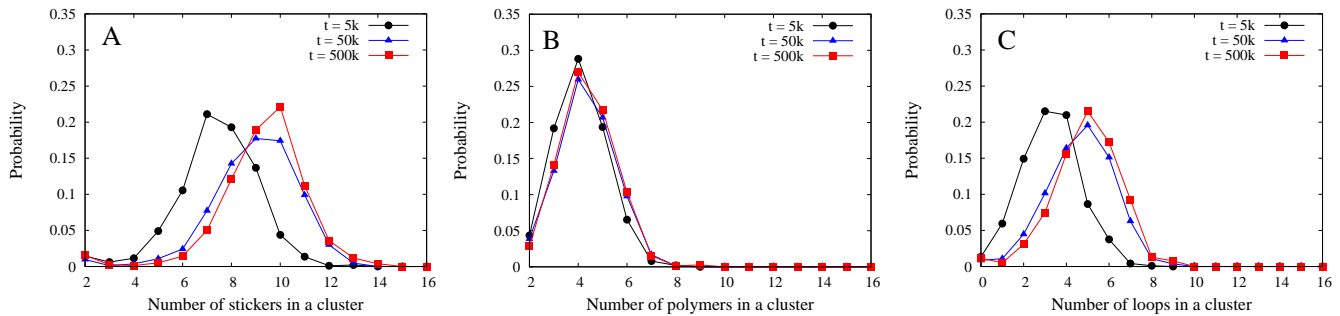}
	\caption{ \label{fig.dist_e6} Distribution of  number of stickers (A),  number of polymers (B) and  number of loops (C) in a cluster is shown at $t=5k, 50k$ and $500k$. }
\end{figure*}

\begin{figure}[h]
\includegraphics[width=0.5\linewidth]{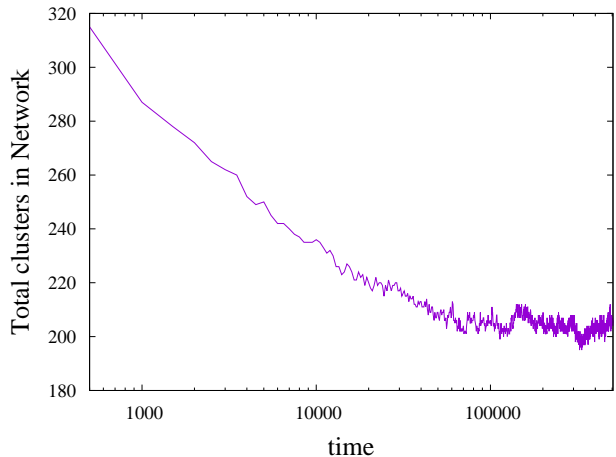}
	\caption{ \label{fig.size} Total number clusters in gel network as a function of time.  }
\end{figure}

In order to study aging we continued our simulations till $t=5\times 10^5 \tau_{LJ}=500k$ ($k \equiv 10^3\tau_{LJ}$). Figure \ref{fig.snapshote6} shows a snapshot (2d projection of a 3d configuration) of all the stickers in the system at time $t=500k$. The non-sticker beads are not shown in the figure. We observe that the stickers form compact clusters of different sizes where the size of a cluster is defined as the number of stickers in it. To quantify the change of cluster size with aging time, we computed the distribution of cluster sizes at three different aging times, $t=5k, 50k$ and $500k$ (see fig. \ref{fig.dist_e6}A). Even though $M_{LC}/M$ saturates already at $t\approx 0.2k$, strong shift of the distribution towards larger cluster sizes was observed as aging time increased from $t=5k$ to $50k$, followed by a more moderate upward shift of the distribution with aging from $t=50k$ to $t=500k$. Since the total number of clusters decreases with time (see fig. \ref{fig.size}) we conclude that the distribution ages due to dissociation of small clusters and the incorporation of the dissociated stickers into larger clusters.
Note that in the latter stages of aging (between $50k$ to $500k$) the shift occurs almost exclusively at the lower end of the distribution and the largest cluster sizes are nearly unaffected by it. This concurs with our intuitive expectation that the size of a cluster is limited from above by packing constraints that arise from excluded volume repulsion between the non-sticker chains that are attached to it. 

In order to obtain further insight into the physical mechanism that leads to the observed aging of the cluster size distribution, we note that each bond between stickers in a cluster is either intramolecular (between stickers of the same polymer) or  intermolecular (between stickers of different polymers) in origin. Thus, a cluster can grow in size either by attachment of stickers of polymers which not yet bound to it, or stickers of polymers that are already bound to that cluster (the latter mechanism leads to the formation of intramolecular  loops). Figures \ref{fig.dist_e6}B and \ref{fig.dist_e6}C show the distribution of number of different polymers and number of loops per cluster measured at three different aging times. Surprisingly, we did not observed any appreciable changes in the distribution of total number of polymers per cluster with aging time whereas the loop distribution shows exactly the same increasing trend with aging as the distribution of the number of stickers per cluster. Since a polymer with $p$ loops that emanate from a cluster contributes $p+1$ stickers to the cluster,  these results suggest that the overall growth in the size of the clusters is the result of progressive loop formation as stickers of polymers that connect larger and smaller clusters, are released in the process of dissociation of the smaller clusters and are reattached to the larger ones, thus further increasing their size.

\begin{figure}[h]
\includegraphics[width=0.5\linewidth]{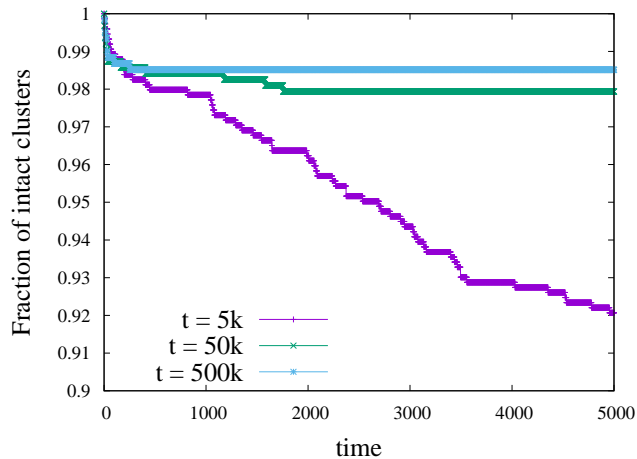}
	\caption{ \label{fig.decay} Plot shows the fraction of intact clusters in the network as a function of aging time. Starting at $t=5k$, approximately $8\%$ of the clusters completely dissociate in $5k$ and the decay continues past this time. As aging time increases, the fraction of dissociated clusters falls to  $1-2\%$ and appears to saturate at this range at later times. }
\end{figure}

Next we examined the effect of aging on the lifetime of clusters in the network. We define the lifetime of a cluster as the time at which the number of the  intermolecular stickers that were {\it {originally}} present in it at time $t_0$, falls below 2 (note that since the distribution of clusters remains fixed in steady state, this means that clusters are being continuously renewed). To measure the average lifetime of a cluster we identified all the clusters in the network containing two or more connected polymers at a particular aging time $t_0$ and monitored their composition (at time intervals of $2 \tau_{LJ}$) for a period of  $\Delta t=5k$ after  $t_0$. The fraction of clusters that were not completely dissolved was computed as a function of time $t$ (see Figure \ref{fig.decay}). Beginning at short aging time, $t_0=5k$, $8\%$ of the initially present clusters have dissolved during $\Delta t=5k$ and continued to decrease after this time ($12\%$ have dissolved by $\Delta t=10k$ - not shown). At longer aging time, $t_0=50k$, only $2\%$ of the initially present clusters have dissolved during $\Delta t=5k$ (less than $4\%$ have dissolved by $\Delta t=10k$). At the longest aging time ($t_0=500k$), only $1\%$ of the clusters have dissolved during $\Delta t=5k$.
Recall that we did not observe any significant changes in the distribution of total number of polymers in the cluster with aging (fig. \ref{fig.dist_e6}B). On the other hand, we defined the stability of a cluster in terms of change of the identity of the polymers attached to that cluster and found that this stability strongly depends on the aging time (fig. \ref{fig.decay}). We conclude that the observed increase of the size of the clusters with aging (see fig. \ref{fig.dist_e6}A) and of their stability are the result of the increase of the number of loops (fig. \ref{fig.dist_e6}C).  Thus, even though the number of polymers in a cluster does not change with time, the number of intermolecular associations between them increases with aging (as each polymer contributes an increasing number of stickers to the cluster) and the binding energy that holds the cluster together increases as the result. Since the dissociation of a sticker from a cluster is a thermally activated process, we expect the detachment time to increase exponentially with the binding energy of the sticker in the cluster \cite{Semenov1995}; therefore, the stability of clusters is expected to dramatically increase with aging, as observed.
\begin{figure}[h]
\includegraphics[width=0.5\linewidth]{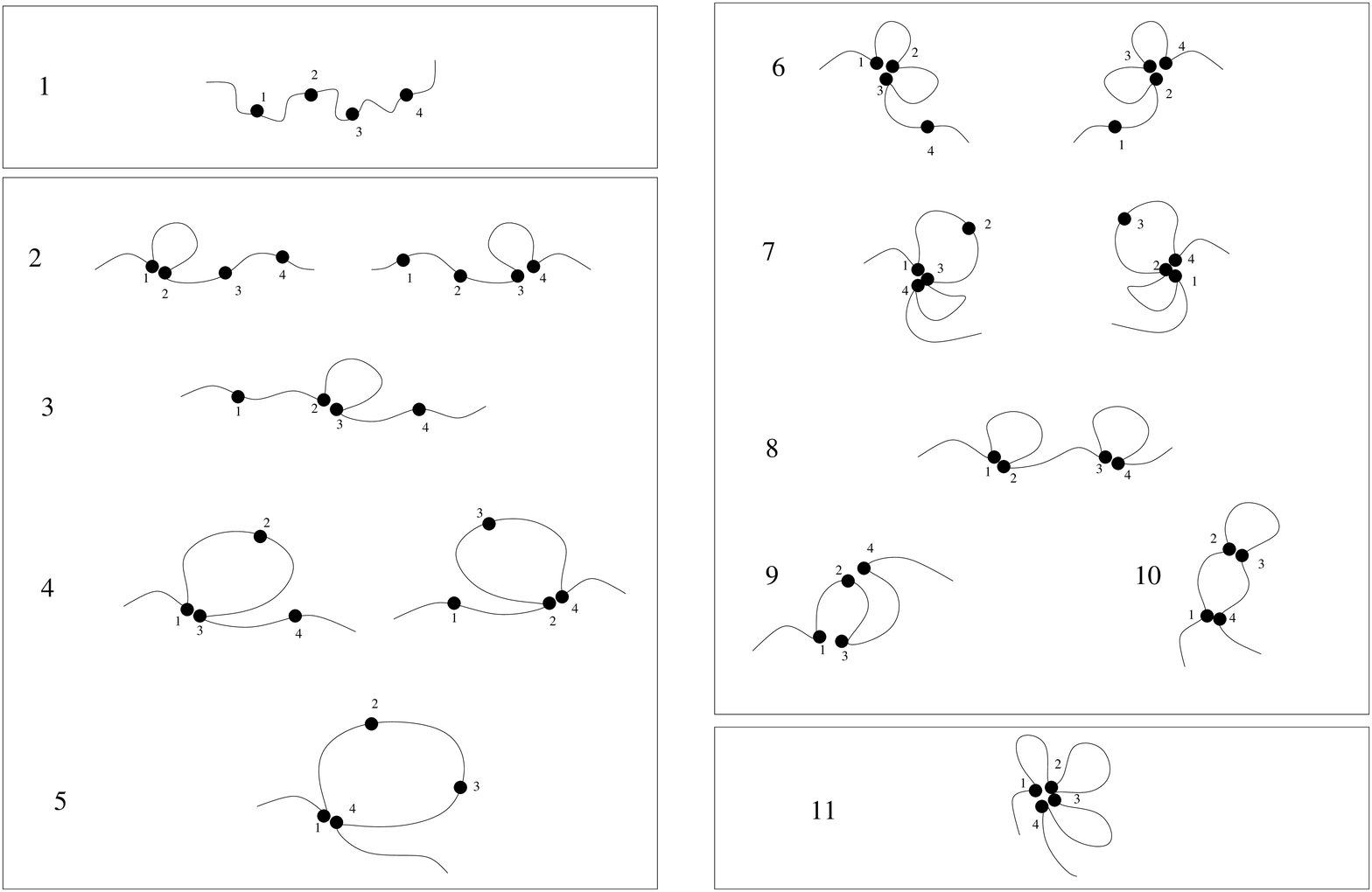}
	\caption{ \label{fig.config} Possible intramolecular associations between stickers (numbered from 1 to 11) of a polymer with four stickers, starting from no associations (configuration 1) to the case when all four stickers belong to the same cluster (configuration 11). }
\end{figure}

\begin{figure*}[ht]
\includegraphics[width=\linewidth]{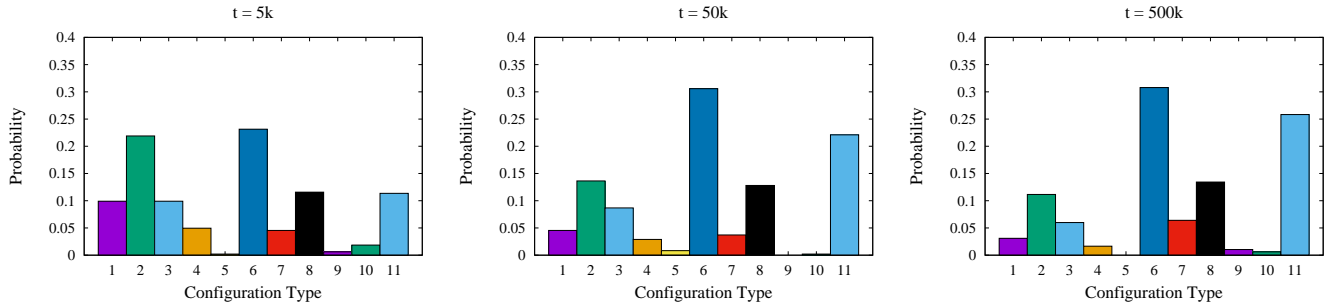}
	\caption{ \label{fig.hist} Histogram of sticker configurations at different aging times. The numbers correspond to configurations shown in Fig. \ref{fig.config}}
\end{figure*}

The configurations of a polymer in a gel can be classified according to the topology of intramolecular associations between its stickers.  Figure \ref{fig.config} shows all the possible configurations of an associating polymer with four stickers.  These configurations range from a fully extended state with no intramolecular associations between its stickers (configuration 1), to a completely collapsed state (configuration 11) where the four stickers form a single aggregate. A histogram that shows the probabilities of occurrence of these configurations in the gel at different aging times is shown in fig. \ref{fig.hist}. As aging time increases the probability of configurations (1-4) with no loops or with a single loop decreases and the probability of configurations (6-8,11) having two and three loops increases. The main change in the histogram takes place between $5k$ and $50k$ and only small aging effects occur later on, a trend similar to that observed in figs. \ref{fig.dist_e6}A and \ref{fig.dist_e6}C and in fig. \ref{fig.decay}.  Configurations 6 and 11 that have the largest intramolecular aggregates (3 and 4 stickers, respectively) become dominant  at longer aging times ($50k$ and $500k$). Chain entropy and excluded volume also play a role in determining the relative probability of different configurations. For example, configuration 7 is suppressed compared to configuration 6 even though they both have an aggregate of 3 stickers, presumably because it is more difficult to accomodate an isolated monomer on a loop in a neighboring intermolecular cluster than a monomer which resides near a free end of a polymer.  

\begin{figure}[h]
\includegraphics[width=0.5\linewidth]{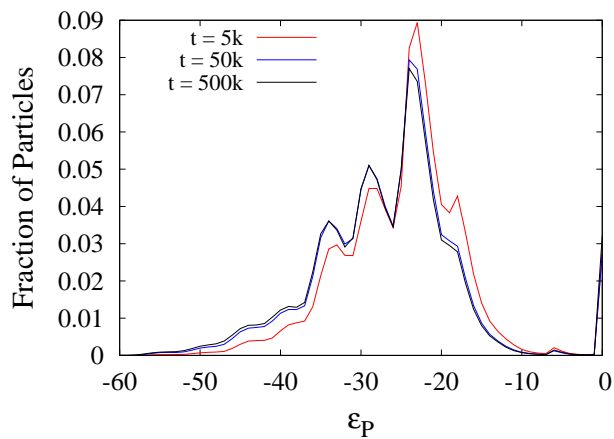}
	\caption{ \label{fig.pe6} Distribution of potential energy of a sticker $\epsilon_P$ at three different aging times}
\end{figure}

The size and geometry of clusters of stickers that crosslink the polymer network is determined by the interplay between binding energy that tends to maximize the number of nearest neighbors of a sticker (e.g., by maximizing the ratio of bulk to surface stickers in the cluster) and by packing constraints due to excluded volume interactions between non-sticker chains attached to the stickers. In order to get some insight about the geometry of the clusters, we computed the potential energy $\epsilon_P$ of a sticker in a cluster by adding the energy of its interactions with all other stickers within cutoff distance ($2.5\sigma$). We carried out this calculation for  all stickers in all clusters and obtained a distribution of sticker energies in the system, which was then time-averaged over a time window of $5k$ after each aging time. Figure \ref{fig.pe6} shows the distribution of potential energy of the stickers at three aging times. Each peak corresponds to a different number of nearest neighbors of the stickers: the peak at $\epsilon_P=0$ corresponds to isolated (non-aggregated) stickers. A second peak at $\epsilon_P\approx -6$ corresponds to stickers that have only one nearest neighbor and its small amplitude reveals that isolated pairs of stickers are rare.  We did not observe a peak that corresponds to two nearest neighbors at  $\epsilon_P\approx -12$. A plausible explanation is that this happens due to the contribution to the potential energy from interactions with second nearest neighbors that have a range of possible radial separations (a plot of the radial distribution is shown in fig S2 in SI) from the sticker,  therefore, a range of interaction energies.  When these interactions are added to the nearest neighbor potential energy contribution and averaged, the second peak is smeared out. To confirm this assertion we computed the average potential energy contribution due to nearest neighbors only (with a cutoff of $1.5\sigma$) and observed a peak corresponding to two nearest neighbors (see fig. S3 in SI). Note that averaging over second nearest neighbor contributions does not effect the peak that corresponds to stickers which have only one nearest neighbor because such stickers mostly form pairs which do not have any second nearest neighbors. The contribution of second nearest neighbors is much smaller than that of nearest neighbors for stickers that have more than two nearest neighbors, and well-resolved peaks are observed for 3, 4, 5 and 6 nearest neighbors ($\epsilon_P\approx -18, -24, -30$ and $-36$, respectively).  A downward shift of the distribution at high potential energies (small number of nearest neighbors) and an upward shift of the distribution at low potential energies (large number of nearest neighbors), is observed at longer aging times. Since increasing the number of nearest neighbors increases the cohesive energy of the cluster and therefore increases its stability, this observation provides additional insights about the connection between the structural reorganization of the network and the enhanced stability of its crosslinks (clusters of stickers), during the aging of the gel. 

\begin{figure*}[ht]
\includegraphics[width=\linewidth]{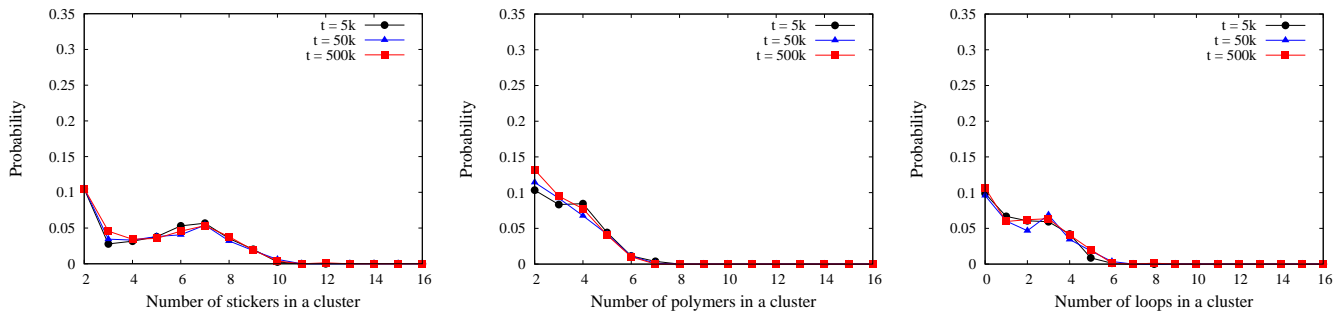}
	\caption{ \label{fig.dist5} Distribution of  number of stickers (A),  number of polymers (B) and  number of loops (C) in a cluster is shown at $t=5k, 50k$ and $500k$, for the $\epsilon_s=5$ system }
\end{figure*}

\begin{figure}[h]
\includegraphics[width=0.5\linewidth]{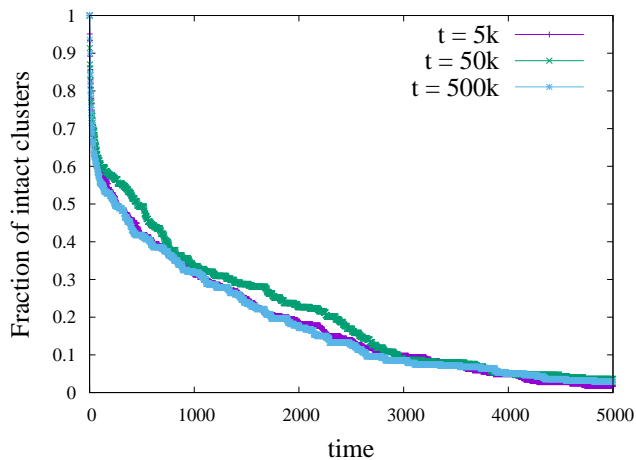}
	\caption{ \label{fig.decaye5} Fraction of intact clusters as a function of time at three different aging times for $\epsilon_s=5$. }
\end{figure}

So far we have explored the aging of a gel made of polymers with strongly associating stickers ($\epsilon_s=6$ in units of $kT$). How sensitive is the observed aging behavior to changes of the interaction strength $\epsilon_s$ between stickers? To explore this question we decreased the interaction parameter from $\epsilon_s=6$ to 
$\epsilon_s=5$ and performed simulations at the same density ($\phi=0.096$) and temperature ($T=1$) as in the $\epsilon_s=6$ case. We first computed the fraction of polymers in the gel and found $M_{LC}/M\approx91\%$, a value that is somewhat lower than $98\%$ found for the $\epsilon_s=6$ system (see fig. S4 in SI). We did not observe appreciable changes of the distributions of number of stickers, polymers and  loops in clusters with aging time in the 
$\epsilon_s=5$ system (see figure  \ref{fig.dist5}). Even though all the distributions were quite broad, they were peaked at 2 stickers, 2 polymers and zero loops per cluster, respectively, from which we concluded that the most probable ``crosslinks'' in the network involve a single association between two polymers. 
The fact that no significant aging was observed in Fig.  \ref{fig.dist5} suggests that unlike the $\epsilon_s=6$ case, for $\epsilon_s=5$ the clusters reach equilibrium on time scale comparable with the time it takes a connected network to form from solution ($M_{LC}/M$ saturates in $0.3k$ - see fig. S4 in SI). Next, we considered the kinetics of dissolution of clusters in the $\epsilon_s=5$ system. No aging is observed in fig. \ref{fig.decaye5} that shows the fraction of intact clusters as a function of time and almost $100\%$ of the clusters dissociated within $\Delta t=5k$ time interval, independent of aging time (recall that only $1\%$ of the clusters dissociated within this time interval in the $\epsilon_s=6$ system upon aging to $500k$). 

\begin{figure}[h]
\includegraphics[width=0.5\linewidth]{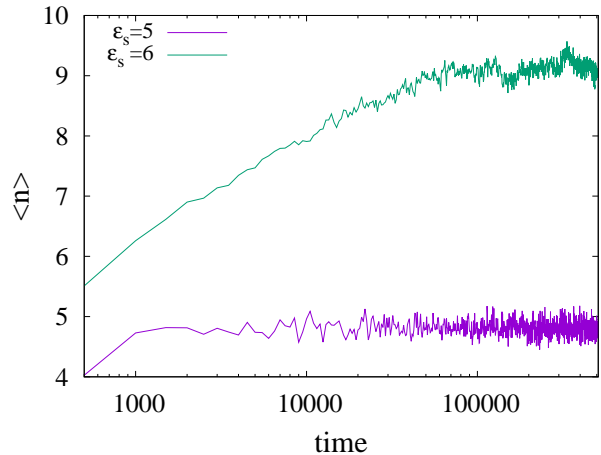}
	\caption{ \label{fig.size5} Average size of clusters as a function of time for $\epsilon_s=5$ and $\epsilon_s=6$ systems. }
\end{figure}

Another comparison of relaxation kinetics in the two systems is shown in Fig. \ref{fig.size5} where we plotted the average number of stickers in a cluster $\langle n(t)\rangle$ as a function of time. We observed that while in the $\epsilon_s=6$ system relaxation to steady state ($\langle n\rangle\approx 9$) takes $100k$ Lennard-Jones time units, in the $\epsilon_s=5$ system steady state is achieved ($\langle n \rangle \le 5$) already after $1k$. 
We also computed the distribution of different polymer configurations in the system and found that for $\epsilon_s=5$ case, the most probable configuration for all aging times is configuration 1 in which there are no loops and therefore no intramolecular associations (see fig. S5  in SI). Since the enhancement of intramolecular associations with time was shown to be the root cause of aging effects in the $\epsilon_s=6$ system, the observation that intramolecular associations are suppressed in the $\epsilon_s=5$ system concurs with the absence of aging in this system.
 
\begin{figure}[h]
\includegraphics[width=0.5\linewidth]{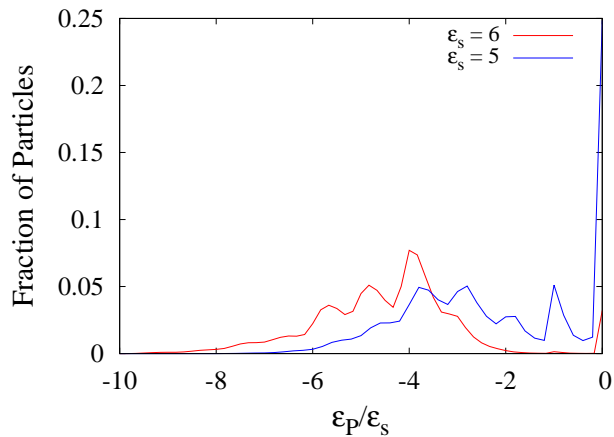}
	\caption{ \label{fig.compare} Distribution of dimensionless potential energy of a sticker $\epsilon_P/\epsilon_s$. The positions of the peaks correspond to the number of nearest neighbors in the cluster (the shift from integer values is due to next nearest neighbor interactions).}
\end{figure}

In order to understand how suppression of intramolecular associations in the $\epsilon_s=5$ system affects the structure of crosslinks (clusters), we calculated the distribution of the potential energy of a sticker $\epsilon_P$ normalized by the value of the interaction parameter, $\epsilon_s$. In the absence of next nearest neighbor interactions one expects to see a distribution with peaks at integers $0, -1, -2, -3, ...$ that correspond to the number of nearest neighbors of a sticker in a cluster (fig. \ref{fig.compare}). For $\epsilon_s=5$ the highest amplitude peaks occur at low values of $\epsilon_P/\epsilon_s$ ($0$ and $-1$), whereas for $\epsilon_s=6$ the highest peaks occur approximately at $-4, -5$ and $-6$. We also computed $\langle \epsilon_P \rangle$ as a function of time for both $\epsilon_s=5$ and $\epsilon_s=6$ cases (see fig. S6 in SI). The magnitude of difference between the steady state values of the average potential energies of $\epsilon_s=5$ and $\epsilon_s=6$ system is $\langle \Delta\epsilon_P \rangle \approx 13$. 
Since the time for thermally activated dissociation of a sticker from a cluster increases exponentially with $\langle \epsilon_P \rangle$ \cite{Semenov1995} (actually this energy is reduced by excluded volume repulsion between the non-sticker segments of the polymer than emanate from the cluster), this explains the dramatic increase of the stability of clusters in the $\epsilon_s=6$ compared to the $\epsilon_s=5$ system (see movies M1 and M2 in SI where a single cluster of six stickers is followed during $\Delta t=5k$ after aging to $t=50k$, for $\epsilon_s=6$ and $\epsilon_s=5$ cases, respectively). 

\section{Discussion}

\begin{figure}[h]
\includegraphics[width=0.5\linewidth]{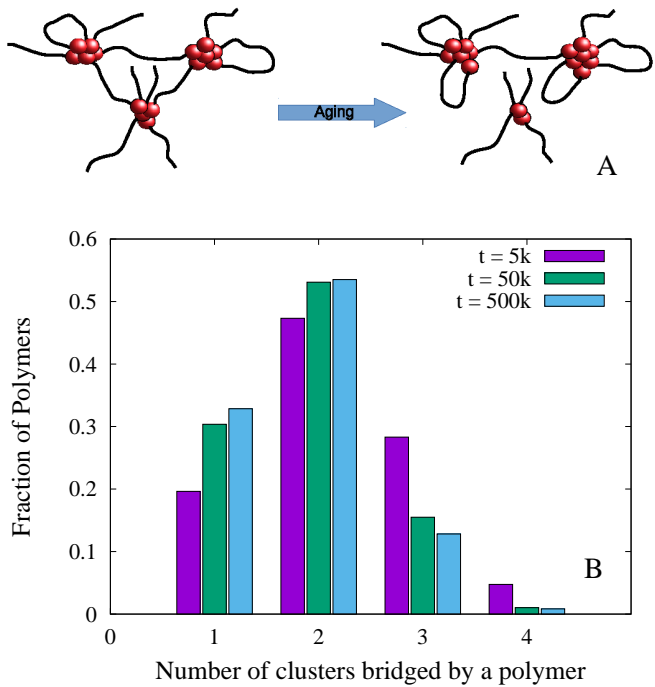}
	\caption{ \label{fig.schematic} (A) Schematic description of the dissolution of a small cluster and the reabsorption of its stickers by neighboring connected clusters via loop formation. (B) Histogram of the number of polymers that bridge between $n$ clusters ($n=1-4$)}
\end{figure}

In this work we studied the aging of thermoreversible physical gels of polymers that contain strongly associating segments (stickers) connected by long soluble chain segments. These polyamphiphilic molecules assemble into a network of connected clusters that act as effective crosslinks. When such a network is rapidly formed from solution (e.g., by a temperature quench), initially it contains many unpaired stickers and a broad non-equilibrium distribution of clusters of different sizes. We find that this distribution relaxes with time, as small clusters disintegrate and their stickers are absorbed by larger ones. The total number of clusters decreases and their size increases with time and this process is accompanied by increase in the number of intramolecular associations/loops (fig. \ref{fig.schematic}A).  In order to understand how the process shown in fig. \ref{fig.schematic}A affects the connectivity of the network, in fig. \ref{fig.schematic}B we present the histogram of the fraction of polymers that take part in one cluster, and those that bridge between $2, 3$ or $4$ different clusters. We observe that the most probable configuration is of a polymer that bridges between $2$ clusters and that aging decreases the number of polymers that bridge between $3$ or $4$ clusters and increases the number of polymers that bridge between $1$ or $2$ clusters. We therefore conclude that while aging increases the size of the clusters and therefore increases their stability, it reduces the bridging between the clusters and therefore decreases the connectivity of the network.

While a detailed study of the viscoelastic behavior of such  thermoreversible gels is beyond the scope of this work we would like to make some qualitative comments about possible effects of aging. Since shear moduli of polymer gels increase with the density of crosslinks \cite{Rubinstein} we expect the high frequency plateau modulus that controls the response of our thermoreversible gels to small deformations, to decrease with aging.  Since the characteristic stress relaxation time is related to the the process of removing a sticker from the cluster,  we expect the frequency at which the response of the gel to shear deformation changes from viscous to elastic, to decrease with aging. The nonlinear elastic response is expected to be controlled by the stability of crosslinks (clusters) and therefore the resistance to large deformations (and fracture) should increase with aging.

Note that in some respects the observed aging in our system resembles glassy behavior in solids - a system out of equilibrium that can only reorganize by overcoming increasingly higher energy barriers \cite{Stein1984}. In our case the barrier corresponds to the energy cost of moving a sticker from one cluster to another and, since the energy barrier increases with cluster size \cite{Semenov1995}, the process slows down with time as the average size of the clusters increases. The analogy is incomplete since, unlike in glasses, in our system the process eventually terminates and further reorganization stops when  clusters reach their equilibrium size for the given sticker pair interaction parameter ($\epsilon_s=6$), temperature ($T=1$) and polymer volume fraction ($\phi=0.096$). For weaker sticker interaction ($\epsilon_s=5$) relaxation to equilibrium is much faster and takes place simultaneously with the formation of the network. For larger interaction parameters we expect aging time to increase dramatically; equilibration would require much longer simulation times and the system may become jammed on accessible time scales. 

We expect the results of this work to be relevant to synthetic physical gels in which the strength of associations may be tuned by changing temperature \cite{Rabin}, pH \cite{Lauber2017}, etc. They may also be relevant to biomolecular aggregates of  intrinsically disordered proteins such as  FG nucleoporins that can form liquid-like microdroplets which remain stable for hours and eventually solidify \cite{Lemke2020}. Aging has also been observed in ameloid-like gels of nucleoporins in which initial aggregation due to hydrophobic interactions between FG repeats was followed by formation of rigid $\beta$ sheet structures that stabilize the network \cite{Gorlich2010}. Note that in our model the intermolecular associations between the polymers become stronger due to intramolecular association between stickers of the same chain because now each polymer contributes several stickers to the cluster, therefore stabilizing it. We propose that  in the case of intrinsically disordered proteins  enhancement of  intramolecular associations and stabilization of clusters during the process of aging corresponds to formation of $\alpha$ helices or $\beta$ sheets and subsequent strengthening of intermolecular aggregates due to strong intermolecular interactions between these elements of secondary structure.

{\em Acknowledgement:} YR would like to acknowledge helpful discussions and correspondence with Eduard Lemke. This work was supported by grants from the Israel Science Foundation and from the Israeli Centers for Research Excellence program of the Planning and Budgeting Committee.

\bibliography{references}

\begin{thebibliography}{21}
\expandafter\ifx\csname natexlab\endcsname\relax\def\natexlab#1{#1}\fi
\expandafter\ifx\csname bibnamefont\endcsname\relax
  \def\bibnamefont#1{#1}\fi
\expandafter\ifx\csname bibfnamefont\endcsname\relax
  \def\bibfnamefont#1{#1}\fi
\expandafter\ifx\csname citenamefont\endcsname\relax
  \def\citenamefont#1{#1}\fi
\expandafter\ifx\csname url\endcsname\relax
  \def\url#1{\texttt{#1}}\fi
\expandafter\ifx\csname urlprefix\endcsname\relax\def\urlprefix{URL }\fi
\providecommand{\bibinfo}[2]{#2}
\providecommand{\eprint}[2][]{\url{#2}}

\bibitem[{\citenamefont{Strauss}(1989)}]{Strauss1989}
\bibinfo{author}{\bibfnamefont{U.~P.} \bibnamefont{Strauss}},
  \bibinfo{journal}{Advances in Chemistry} \textbf{\bibinfo{volume}{223}},
  \bibinfo{pages}{317} (\bibinfo{year}{1989}).

\bibitem[{\citenamefont{Chassenieux et~al.}(2011)\citenamefont{Chassenieux,
  Nicolai, and Benyahia}}]{Chassenieux2011}
\bibinfo{author}{\bibfnamefont{C.}~\bibnamefont{Chassenieux}},
  \bibinfo{author}{\bibfnamefont{T.}~\bibnamefont{Nicolai}}, \bibnamefont{and}
  \bibinfo{author}{\bibfnamefont{L.}~\bibnamefont{Benyahia}},
  \bibinfo{journal}{Current Opinion in Colloid and Interface Science}
  \textbf{\bibinfo{volume}{16}}, \bibinfo{pages}{18} (\bibinfo{year}{2011}).

\bibitem[{\citenamefont{Zhang et~al.}(2018)\citenamefont{Zhang, Chen, and
  Colby}}]{Colby}
\bibinfo{author}{\bibfnamefont{Z.}~\bibnamefont{Zhang}},
  \bibinfo{author}{\bibfnamefont{Q.}~\bibnamefont{Chen}}, \bibnamefont{and}
  \bibinfo{author}{\bibfnamefont{R.~H.} \bibnamefont{Colby}},
  \bibinfo{journal}{Soft Matter} \textbf{\bibinfo{volume}{14}},
  \bibinfo{pages}{2961} (\bibinfo{year}{2018}).

\bibitem[{\citenamefont{Cates and Witten}(1986)}]{Cates1986}
\bibinfo{author}{\bibfnamefont{M.~E.} \bibnamefont{Cates}} \bibnamefont{and}
  \bibinfo{author}{\bibfnamefont{T.~A.} \bibnamefont{Witten}},
  \bibinfo{journal}{Macromolecules} \textbf{\bibinfo{volume}{19}},
  \bibinfo{pages}{732} (\bibinfo{year}{1986}).

\bibitem[{\citenamefont{Borisov and Halperin}(1995)}]{Borisov1995}
\bibinfo{author}{\bibfnamefont{O.~V.} \bibnamefont{Borisov}} \bibnamefont{and}
  \bibinfo{author}{\bibfnamefont{A.}~\bibnamefont{Halperin}},
  \bibinfo{journal}{Langmuir} \textbf{\bibinfo{volume}{11}},
  \bibinfo{pages}{2911} (\bibinfo{year}{1995}).

\bibitem[{\citenamefont{Semenov et~al.}(1995)\citenamefont{Semenov, Joanny, and
  Khokhlov}}]{Semenov1995}
\bibinfo{author}{\bibfnamefont{A.~N.} \bibnamefont{Semenov}},
  \bibinfo{author}{\bibfnamefont{J.-F.} \bibnamefont{Joanny}},
  \bibnamefont{and} \bibinfo{author}{\bibfnamefont{A.~R.}
  \bibnamefont{Khokhlov}}, \bibinfo{journal}{Macromolecules}
  \textbf{\bibinfo{volume}{28}}, \bibinfo{pages}{1066—1075}
  (\bibinfo{year}{1995}).

\bibitem[{\citenamefont{Winnik and Yektaf}(1997)}]{Winnik1997}
\bibinfo{author}{\bibfnamefont{M.~A.} \bibnamefont{Winnik}} \bibnamefont{and}
  \bibinfo{author}{\bibfnamefont{A.}~\bibnamefont{Yektaf}},
  \bibinfo{journal}{Current Opinion in Colloid and Interface Science}
  \textbf{\bibinfo{volume}{2}}, \bibinfo{pages}{424} (\bibinfo{year}{1997}).

\bibitem[{\citenamefont{Semenov and Rubinstein}(1998)}]{Semenov1}
\bibinfo{author}{\bibfnamefont{A.~N.} \bibnamefont{Semenov}} \bibnamefont{and}
  \bibinfo{author}{\bibfnamefont{M.}~\bibnamefont{Rubinstein}},
  \bibinfo{journal}{Macromolecules} \textbf{\bibinfo{volume}{31}},
  \bibinfo{pages}{1373} (\bibinfo{year}{1998}).

\bibitem[{\citenamefont{Rubinstein and Semenov}(1998)}]{Semenov2}
\bibinfo{author}{\bibfnamefont{M.}~\bibnamefont{Rubinstein}} \bibnamefont{and}
  \bibinfo{author}{\bibfnamefont{A.~N.} \bibnamefont{Semenov}},
  \bibinfo{journal}{Macromolecules} \textbf{\bibinfo{volume}{31}},
  \bibinfo{pages}{1386} (\bibinfo{year}{1998}).

\bibitem[{\citenamefont{Rubinstein and Dobrynin}(1999)}]{Rubinstein1999}
\bibinfo{author}{\bibfnamefont{M.}~\bibnamefont{Rubinstein}} \bibnamefont{and}
  \bibinfo{author}{\bibfnamefont{A.~V.} \bibnamefont{Dobrynin}},
  \bibinfo{journal}{Current Opinion in Colloid and Interface Science}
  \textbf{\bibinfo{volume}{4}}, \bibinfo{pages}{83 } (\bibinfo{year}{1999}).

\bibitem[{\citenamefont{Dobrynin}(2004)}]{Dobrynin2004}
\bibinfo{author}{\bibfnamefont{A.~V.} \bibnamefont{Dobrynin}},
  \bibinfo{journal}{Macromolecules} \textbf{\bibinfo{volume}{37}},
  \bibinfo{pages}{3881} (\bibinfo{year}{2004}).

\bibitem[{\citenamefont{Rubinstein and Colby}(2003)}]{Rubinstein}
\bibinfo{author}{\bibfnamefont{M.}~\bibnamefont{Rubinstein}} \bibnamefont{and}
  \bibinfo{author}{\bibfnamefont{R.~H.} \bibnamefont{Colby}},
  \emph{\bibinfo{title}{Polymer Physics}} (\bibinfo{publisher}{Oxford
  University Press}, \bibinfo{year}{2003}).

\bibitem[{\citenamefont{Gomez-Solano et~al.}(2013)\citenamefont{Gomez-Solano,
  Blickle, and Bechinger}}]{Gomez2013}
\bibinfo{author}{\bibfnamefont{J.~R.} \bibnamefont{Gomez-Solano}},
  \bibinfo{author}{\bibfnamefont{V.}~\bibnamefont{Blickle}}, \bibnamefont{and}
  \bibinfo{author}{\bibfnamefont{C.}~\bibnamefont{Bechinger}},
  \bibinfo{journal}{Physical Review E} \textbf{\bibinfo{volume}{87}},
  \bibinfo{pages}{012308} (\bibinfo{year}{2013}).

\bibitem[{\citenamefont{Secchi et~al.}(2013)\citenamefont{Secchi, Roversi,
  Buzzaccaro, Piazza, and Piazza}}]{Piazza2013}
\bibinfo{author}{\bibfnamefont{E.}~\bibnamefont{Secchi}},
  \bibinfo{author}{\bibfnamefont{T.}~\bibnamefont{Roversi}},
  \bibinfo{author}{\bibfnamefont{S.}~\bibnamefont{Buzzaccaro}},
  \bibinfo{author}{\bibfnamefont{L.}~\bibnamefont{Piazza}}, \bibnamefont{and}
  \bibinfo{author}{\bibfnamefont{R.}~\bibnamefont{Piazza}},
  \bibinfo{journal}{Soft Matter} \textbf{\bibinfo{volume}{9}},
  \bibinfo{pages}{3931} (\bibinfo{year}{2013}).

\bibitem[{\citenamefont{Schupper et~al.}(2008)\citenamefont{Schupper, Rabin,
  and Rosenbluh}}]{Rabin}
\bibinfo{author}{\bibfnamefont{N.}~\bibnamefont{Schupper}},
  \bibinfo{author}{\bibfnamefont{Y.}~\bibnamefont{Rabin}}, \bibnamefont{and}
  \bibinfo{author}{\bibfnamefont{M.}~\bibnamefont{Rosenbluh}},
  \bibinfo{journal}{Macromolecules} \textbf{\bibinfo{volume}{41}},
  \bibinfo{pages}{3983} (\bibinfo{year}{2008}).

\bibitem[{\citenamefont{Osmanovic and Rabin}(2018)}]{Dino}
\bibinfo{author}{\bibfnamefont{D.}~\bibnamefont{Osmanovic}} \bibnamefont{and}
  \bibinfo{author}{\bibfnamefont{Y.}~\bibnamefont{Rabin}},
  \bibinfo{journal}{Biophysical Journal} \textbf{\bibinfo{volume}{114}},
  \bibinfo{pages}{534 } (\bibinfo{year}{2018}).

\bibitem[{\citenamefont{Plimpton}(1995)}]{Plimpton1995}
\bibinfo{author}{\bibfnamefont{S.}~\bibnamefont{Plimpton}},
  \bibinfo{journal}{Journal of Computational Physics}
  \textbf{\bibinfo{volume}{117}}, \bibinfo{pages}{1 } (\bibinfo{year}{1995}).

\bibitem[{\citenamefont{Palmer et~al.}(1984)\citenamefont{Palmer, Stein,
  Abrahams, and Anderson}}]{Stein1984}
\bibinfo{author}{\bibfnamefont{R.~G.} \bibnamefont{Palmer}},
  \bibinfo{author}{\bibfnamefont{D.~L.} \bibnamefont{Stein}},
  \bibinfo{author}{\bibfnamefont{E.}~\bibnamefont{Abrahams}}, \bibnamefont{and}
  \bibinfo{author}{\bibfnamefont{P.~W.} \bibnamefont{Anderson}},
  \bibinfo{journal}{Phys. Rev. Lett.} \textbf{\bibinfo{volume}{54}},
  \bibinfo{pages}{1965} (\bibinfo{year}{1984}).

\bibitem[{\citenamefont{Lauber et~al.}(2017)\citenamefont{Lauber, Depoorter,
  Nicolai, Chassenieux, and Colombani}}]{Lauber2017}
\bibinfo{author}{\bibfnamefont{L.}~\bibnamefont{Lauber}},
  \bibinfo{author}{\bibfnamefont{J.}~\bibnamefont{Depoorter}},
  \bibinfo{author}{\bibfnamefont{T.}~\bibnamefont{Nicolai}},
  \bibinfo{author}{\bibfnamefont{C.}~\bibnamefont{Chassenieux}},
  \bibnamefont{and}
  \bibinfo{author}{\bibfnamefont{O.}~\bibnamefont{Colombani}},
  \bibinfo{journal}{Macromolecules} \textbf{\bibinfo{volume}{50}},
  \bibinfo{pages}{8178} (\bibinfo{year}{2017}).

\bibitem[{\citenamefont{Celetti et~al.}(2020)\citenamefont{Celetti, Paci,
  Caria, VanDelinder, Bachand, and Lemke}}]{Lemke2020}
\bibinfo{author}{\bibfnamefont{G.}~\bibnamefont{Celetti}},
  \bibinfo{author}{\bibfnamefont{G.}~\bibnamefont{Paci}},
  \bibinfo{author}{\bibfnamefont{J.}~\bibnamefont{Caria}},
  \bibinfo{author}{\bibfnamefont{V.}~\bibnamefont{VanDelinder}},
  \bibinfo{author}{\bibfnamefont{G.}~\bibnamefont{Bachand}}, \bibnamefont{and}
  \bibinfo{author}{\bibfnamefont{E.~A.} \bibnamefont{Lemke}},
  \bibinfo{journal}{J.Cell Biol.} \textbf{\bibinfo{volume}{219}},
  \bibinfo{pages}{1} (\bibinfo{year}{2020}).

\bibitem[{\citenamefont{Adera et~al.}(2010)\citenamefont{Adera, Frey, Maasc,
  Schmidt, Görlich, and Baldus}}]{Gorlich2010}
\bibinfo{author}{\bibfnamefont{C.}~\bibnamefont{Adera}},
  \bibinfo{author}{\bibfnamefont{S.}~\bibnamefont{Frey}},
  \bibinfo{author}{\bibfnamefont{W.}~\bibnamefont{Maasc}},
  \bibinfo{author}{\bibfnamefont{H.~B.} \bibnamefont{Schmidt}},
  \bibinfo{author}{\bibfnamefont{D.}~\bibnamefont{Görlich}}, \bibnamefont{and}
  \bibinfo{author}{\bibfnamefont{M.}~\bibnamefont{Baldus}},
  \bibinfo{journal}{PNAS} \textbf{\bibinfo{volume}{107}},
  \bibinfo{pages}{6281–6285} (\bibinfo{year}{2010}).

\end{thebibliography}

\end{document}